\documentstyle[aps,twocolumn]{revtex}
\begin{document}
\title{Electromigration-Induced Flow of Islands and Voids \\
on the Cu(001) Surface}
\author{Hanoch Mehl, Ofer Biham and Oded Millo}
\address{
Racah Institute of Physics,
The Hebrew University,
Jerusalem 91904,
Israel}
\author{Majid Karimi}
\address{
Physics Department,
Indiana University of Pennsylvania,
Indiana, PA 15705
}
\maketitle

\begin{abstract}
Electromigration-induced flow of islands and voids on
the Cu(001) surface is studied at the atomic scale.
The basic drift mechanisms are
identified using a complete set of energy barriers for
adatom hopping on the Cu(001) surface,
combined with kinetic Monte Carlo simulations. 
The energy barriers are calculated by the embedded atom method, 
and parameterized using a simple model.
The dependence of the flow on the temperature,
the size of the clusters, and the strength
of the applied field is obtained.
For both islands and voids it is found that
edge diffusion is the dominant mass-transport
mechanism. The rate limiting steps are identified.
For both islands and voids they involve detachment of
atoms from corners into the adjacent edge.
The energy barriers for these moves are found to be
in good agreement with the activation energy for 
island/void drift obtained from Arrhenius analysis
of the simulation results.
The relevance of the results to other FCC(001)
metal surfaces and their 
experimental implications are discussed.

\end{abstract}


\section{Introduction}

Electromigration (EM) describes the biased
diffusion processes of bulk and surface
atoms under the influence of an applied electric field
\cite{Huntington1975,Ho1989,Vook1995}. 
The driving force is composed of two major components:
one is the direct electrostatic interaction between the
applied field and the conductor ions and the second, denoted
as the ``wind force'' is due to momentum transfer from the
conduction electrons impinging upon the ionic cores.
The EM force 
$F$
is characterized by an effective charge
$Z^*$
of the ions, which quantifies its strength according to
$F= e Z^* E$,
where
$E$ is the electric field.
In metals, the wind force is the dominant component, thus
$Z^*$ is negative \cite{Ho1989,Rous1994}.
The problem of EM in metal films has been extensively 
investigated experimentally and theoretically for over
three decades
\cite{Black1967,Blech1967,Ho1966,Rosenberg1968}.
The interest in this problem was motivated 
by the fact that EM has been
identified as a major failure mode of metal interconnects
in microelectronic devices, where current densities
as large as $10^6\mbox{ A/cm}^2$, are typical
\cite{Lloyd1995,Filter1996}.
Indeed, significant 
experimental effort was directed at material
and technological issues, 
aimed to reduce EM-induced damage.
These issues include the choice of metal to be used
(e.g. Al, AlCu alloys, or Cu dominated alloys), 
deposition procedures and wire geometries. 

Statistical properties of EM phenomena have
been studied extensively as a 
function of applied current density and temperature.
These properties include the
time to failure of metal wires 
under current stress
\cite{Ho1989,Black1967,Gladkikh1996},
drift velocities
\cite{Hu1996,Gall1996},
resistance-noise
\cite{Dagge1996},
as well as 
resistance increase rates 
\cite{Vook1995,Gladkikh1996,Jo1995a,Verb1996}.
Experimental results were 
analyzed using phenomenological
equations, such as the Black equation
\cite{Black1967}, 
that relates the mean
time 
to failure 
$t_{mean}$
to a 
{\it single} activation energy $Q$ through
$t_{mean} \propto j^{-2} \exp{(Q/k_B T)}$,
where
$j$ is the current density and $T$ is the temperature.
A similar equation is also used to analyze 
the resistance-increase rate
\cite{Vook1995,Park1991,Vook1996}.
This approach provides a useful characterization of the wire
performance and reliability.
However, it does not yield much insight into the fundamental 
processes that give rise to EM.
Particularly, the assumption that there is only one
activation energy is inadequate.
This is manifested in the dispersion of activation energies
reported in the literature for a given material.
Taking Cu as an important example, 
we find a broad range of reliably measured activation energies, e.g.,  
0.47 eV 
\cite{Jo1995a},
0.7-0.9 eV
\cite{Hu1996},
0.79 eV
\cite{Park1991}
and
1.21 eV
\cite{Gladkikh1996}.
These results indicate that EM is not 
driven by a single atomic diffusion 
process,
but is a complex phenomenon
that involves a wide spectrum of activation energies,
and depends on the microscopic details of the samples.

A large variety of EM-related phenomena have been observed,
such as thinning, void (and hillock) formation, 
followed by their growth and migration
\cite{Ho1989,Vook1995}.  
The rate at which each of these processes develops
depends on the dominant atomic 
diffusion mechanism involved.
In particular, one can distinguish between atom diffusion
inside crystallites, along grain 
boundaries, and diffusion on surfaces and interfaces.
Grain boundary diffusion is believed 
to be dominant in polycrystalline materials
\cite{Ho1989,Vook1995}.
However, surface (and interface)
diffusion becomes more important as the cross section
of the wire decreases.
Hence, studies of surface EM at the atomic and nanometer scales
may provide much insight into the
fundamental EM-induced processes.
Recent studies of surface EM on single crystal Si samples
have focused on EM-induced step dynamics
\cite{Williams1995a,Yang1996,Fu1997}. 
The mobility of steps and dislocations in
thin polycrystalline metal films 
has also been studied
\cite{Shimoni1997,Shimoni1998}.
In theoretical studies 
using a continuum description, the effects of EM on atomic steps 
\cite{Kandel1995}
and voids 
\cite{Kraft1996,Schimschak1998} 
were examined.  

In this paper we present a detailed study of diffusion of single
islands and voids of monoatomic height
on the Cu(001) surface, under electromigration conditions.
In our analysis we employ  a complete set of hopping energy
barriers for Cu atoms on the Cu(001) surface, obtained using 
the embedded atom method (EAM)
\cite{Daw1983}. 
The biased diffusion of islands and voids is studied using
energy considerations and kinetic Monte-Carlo (MC)
simulations.
The different mechanisms involved in the diffusion of islands
and voids are identified and 
the activation energies of the corresponding
rate-limiting steps are quantified. 
The drift velocities are obtained as a 
function of the EM bias, island/void 
size and the surface temperature. 
For both islands and voids the slopes of the Arrhenius
plots obtained from the simulation results
are found to be
in excellent agreement with the activation energies
of the rate-limiting steps, and thus confirm
their identification. 
The drift velocities are found to be linear with
the bias magnitude. 

The paper is organized as follows. In Sec. II we present the
model and assumptions. 
The relevant atomic processes are presented in Sec. III.
The simulations and results are shown
in Sec. IV, followed by a discussion and summary
in Sec. V.

\section{Model and Assumptions}

Surface diffusion is described by thermally activated
hopping processes of adatoms.
The hopping rate $h$ (in units of hops per second)
of a given atom to each unoccupied nearest neighbor (NN)
site is given by 

\begin{equation}
h=\nu \cdot \exp(-E_B/k_BT)
\label{hoppingrate}
\end{equation}

\noindent
where 
$\nu = 10^{12}$ $s^{-1}$ 
is the commonly used 
attempt rate,
$E_B$ is the activation energy barrier, 
$k_B$ is the Boltzmann constant and $T$ is the temperature.

The activation energy barrier $E_B$ depends 
on the local environment of the
hopping atom, namely the configuration of occupied and
unoccupied adjacent sites.
We assume that only nearest and next-nearest 
neighbor sites have
a non-negligible effect on the activation energy.
On FCC(001) surfaces, under these assumptions,
the hopping energy barrier is
determined by the occupation of
seven adjacent sites, as shown in Fig. \ref{fig:env}. 
In this model, 
adatoms diffuse along the
principal crystal directions. 
On the FCC(001) surface,
the possible paths are
along the $x=<011>$ and $y=<01\bar{1}>$ axes.

In the simulations described below we use
a set of energy barriers for Cu on Cu(001) 
\cite{Karimi1995,Biham1998}, which includes all
possible local environments of the hoping atom.
The effect of the electric current is 
incorporated by lowering the barrier for
diffusion in the direction of the electron flow, 
and by raising it in the opposite
direction. 
Thus, if the barrier for a certain process is
$E_B$ 
without bias, it will be $E_B-\Delta$ for that process in the direction
of the bias, and  $E_B+\Delta$ in the opposite direction.
The bias $\Delta$ is proportional to the applied field
according to $\Delta=eZ^*E\cdot a$, where $a$ is the projection
of the hopping distance along the bias direction.
In this paper we focus on the case in which the field 
is parallel to either the $x$ or the $y$ axis. Thus, 
for Cu, $a=2.55\AA$, which is the lattice constant of
the resulting two dimensional square lattice.
We also assume that  
bias $\Delta$ is the same for all processes and barriers.
Although the distortion in the energy landscape may depend on the local
configuration, 
no details are known about this dependence.
It is also assumed that the dependence of 
the attempt frequency
$\nu$
on the field is weak enough, and can be neglected.

Consider a single object 
(adatom, dimer, vacancy etc.) 
diffusing on the Cu(001) surface.
The net drift of this object 
in the $x$ direction 
after time $t$,
will be:
$\Delta x=n_+-n_-$,  where $n_+$ and $n_-$ 
are the numbers of hops in the positive
and negative directions of the $x$ axis respectively.
On time scales much longer than the time for a single hop, the average
number of hops will be proportional to the 
elapsed time and to the hopping rate. 
Assume now that the bias is in the positive $x$  direction.
According to the model we have:

\begin{equation}
<n_{\pm}>=t\nu \cdot e^{-\frac{E_B\mp\Delta}{k_BT}},
\end{equation}

\noindent
thus the average drift velocity in the $x$ direction will be:

\begin{equation}
\label{eq:velocity}
<v>=\frac{<x>}{t}=2h\cdot\sinh(\frac{\Delta}{k_BT}),
\end{equation}

\noindent
where $h=\nu\cdot\exp(-E_B/k_BT)$ 
is the rate of the hopping process without any
bias [Eq. (\ref{hoppingrate})]. The drift
velocity of a given object is 
proportional to its hopping rate, namely,
it depends exponentially
on the activation energy, just like the diffusion 
coefficient.
The proportionality factor, $\sinh(\Delta/k_BT)$, 
is common to all processes, and depends on the
temperature and the applied field.
Under physically relevant condition $\Delta\ll k_BT$
(see below). Therefore $\sinh(\Delta/k_BT)$
can be approximated by $\Delta/k_BT$,
and the drift velocity is linearly proportional
to the bias.

\section{Atomic Description of Cluster Drift}

\subsection{The Energy Barriers}

Within the assumptions described above, adatom
diffusion on the Cu(001) surface involves $2^7=128$
hopping processes (Fig. \ref{fig:env}). 
In order to obtain the rates of these
processes one needs to calculate their
energy barriers. Our analysis is based on semi-empirical 
calculations obtained via the 
embedded atom method (EAM)
\cite{Daw1983}. 
This method  
provides a good 
description of
self diffusion of Cu on Cu(001)
\cite{Karimi1995},
as well as self diffusion on other metal surfaces
\cite{Liu1991,Perkins1995}.
Specifically, we use the EAM
functions of Cu developed by Adams, Foiles, and Wolfer
\cite{Adams1989}
which
are fitted to a similar data base as the one employed by Foiles,
Baskes, and Daw
\cite{Foiles1986}.
Generally, these energy barriers are in agreement
with those obtained using other semi-empirical methods
such as the effective medium theory (EMT)
\cite{Jacobsen1987,Hansen1991},
and ab-initio calculations
\cite{Boisvert1997}.
Though each process has a different activation energy,
they can be roughly divided into four groups
\cite{Biham1998,Mehl1999}. 
These groups give rise to four typical time scales in the 
problem which span over several
orders of magnitude.
These time scales correspond (in ascending order) 
to attachment processes, edge diffusion,
free surface hopping and detachment processes.
We will argue that all four scales are essential in 
the description of cluster drift.
Furthermore, even much smaller differences between processes 
which belong to the same group will appear to be significant.

\subsection{Simple Model for Energy Barriers}

The calculated energy barriers can be well approximated
by a simple model which has four parameters
\cite{Biham1998}.
This model provides a systematic description of the
energetics of the hopping processes, and a classification
of these processes to four groups.
It also enables generalization to different materials.
According to the model, the energy barrier $E_B$ for
a certain process depends on the seven sites of
Fig. \ref{fig:env} as follows:

\begin{eqnarray}
E_{B} & = & E_0 + E_{NN}^{in} \cdot (S_3+S_1+S_5) \nonumber \\
      &   &	-E_{NN}^{top} \cdot (S_1+S_5+S_2+ S_6) \nonumber \\
      &   &	+ E_{NNN} \cdot (S_0 + S_2 + S_4 + S_6) 
\label{barrier4}
\end{eqnarray}
where $S_i=1$ if site $i$ is occupied and $S_i=0$ if it is vacant,
$E_0$ is the barrier for an isolated adatom hopping on 
the surface, $E_{NN}^{in}$ and $E_{NNN}$ are the effective binding
energies of the hopping atom to its nearest and 
next-nearest neighbors, respectively.
The highest energy along the hopping path is obtained
in the vicinity of the bridge site. The binding energy
of the hopping adatom at this highest point, to an atom in one
of the four adjacent sites is given by $E_{NN}^{top}$.
Here we use a simplified version of the model 
assuming that $E_{NN}^{in}\simeq E_{NN}^{top}$. This
approximate degeneracy is found for most FCC metals 
\cite{Biham1998,Mehl1999}.
The expression we use for the barriers is thus
\begin{eqnarray}
E_{B}^n & = & E_0 + E_{NN} \cdot (S_3 - S_2 - S_6) \nonumber \\
        &   &	+ E_{NNN} \cdot (S_0 + S_2 + S_4 + S_6). 
\label{eq:model}
\end{eqnarray}

\noindent The three parameters $E_0$, $E_{NN}$ and $E_{NNN}$
were estimated for Cu(001) 
and yielded $E_0=0.49$ eV, $E_{NN}=0.27$ eV and 
$E_{NNN}=0.027$ eV
\cite{Biham1998,Mehl1999}.
While simple and intuitive, the barriers obtained from
the model may deviate in some cases significantly from
the EAM barriers. Such deviations occur in 
relatively dense local environments, where several
sites in both sides of the hopping atom are occupied.
The model accounts for the complex interactions between 
these atoms only on the average.
In the following discussion barriers obtained by both
EAM and the model of Eq. (\ref{eq:model})
are quoted. The data for the simulation results presented 
below were obtained using the EAM barriers.

\subsection{Physical Conditions}

The following discussion concentrates on clusters
of 60-1000 atoms/vacancies on the Cu(001) surface.
Such clusters are typically created during deposition
and sputtering experiments
\cite{Pai1997,Stoldt1998}. 
We consider temperatures in the range 200-600K, 
which is the relevant range for most experimental
studies. The model for 
diffusion described above applies well in this regime,
where no other hopping mechanisms are significant.
At higher temperatures additional processes may take place
\cite{Boisvert1997}.

The effective charge for EM in bulk copper was found experimentally to be
$Z^*\approx -5$
\cite{Huntington1975}.
For surface EM there exist only theoretical estimations. These calculations 
found the effective charge
to be of the same order of magnitude, 
$Z^*\approx -20$
\cite{Rous1994,Rous1996}, 
almost independent of surface orientation and diffusion path.
Typical current densities in EM experiments (and in integrated circuits) 
are $j\approx10^6-10^7\mbox{A/cm}^2$. 
The resistivity of Cu under typical conditions is
$\rho\approx 2\cdot 10^{-6} \Omega \cdot$cm.
If we assume that the current is distributed uniformly
across the conductor's cross-section, we find the EM force acting on surface
atom to be 
$F=e Z^* j\rho\approx10$ eV/cm.
For Cu(001) the distance between adjacent sites is $\approx2.55\AA$.
The work done by the EM force during one hop
in the bias direction is thus $\approx 10^{-7}$eV. 
Intuitively, this would be the change in the energy barrier for the process.
Since most of the relevant barriers are in the range 0.1-0.8 eV, 
we find that $\Delta$ is typically 6-7 orders of 
magnitude smaller than $E_B$. 
For such a small bias to produce any effect an
extremely long time is required. 
In experiments, it takes at least several hours at
temperatures higher than $400^\circ C$ for any significant change. Simulating
systems at these temperatures for more than a second is beyond the
power of contemporary computers.
To overcome this difficulty, we use bias values 
in the range $\Delta\approx 10^{-3}-10^{-4}$ eV.
These values are larger than the realistic ones, 
but still much smaller (2-3 orders of magnitude) than
the diffusion barriers. The linear response approximation is thus still
valid.

\subsection{Island Drift}

During the drift, clusters (islands and voids) maintain
an approximate square shape with small fluctuations and
rounded corners. There is experimental evidence
that this is the equilibrium shape of islands
\cite{Pai1997,Stoldt1998}. 
These experiments show that even if a different shape
is created (during coalescence, for example), it rearranges
to the square-like pattern within
several minutes at room temperature.
This is a consequence of the square symmetry of the lattice
and the fast edge diffusion.
This fact has important implication on the atomic
details of island diffusion, with or without bias.
When clusters diffuse, they typically move one lattice
site at a time, while maintaining their equilibrium shape
in the new position. Although the bias drives the
system out of equilibrium,
since the applied bias is small, the 
basic pattern is preserved. 
This feature is also observed in
our computer simulations.
However, in the presence of bias, atoms drift along island
edges with a preferable direction, 
and thus sharpen the 
rounded corners in the bias direction. 
The basic cycle of island drift, in which the center of mass
moves one lattice site in the bias direction, can be divided 
to three main stages.
These are shown schematically in Fig. \ref{fig:island_drift}.
To describe the basic cycle, 
take the starting point to be an island with 
a straight edge in the direction of the drift.
[Fig. \ref{fig:island_drift}(a)]. 
First, an atom detaches from one of the four corners
and arrives at the island front 
[Fig. \ref{fig:island_drift}(b)]. 
Next, several other
atoms nucleate to this atom from the corners
and create a new row of atoms in the front
[Fig. \ref{fig:island_drift}(c)]. Last, atoms from the rear
side of the island fill the place of the atoms that formed
the new row and straighten the corners to 
form a configuration similar to that of 
Fig. \ref{fig:island_drift}(a), shifted by 
one lattice site in the bias direction. 
The entire process can
now repeat as the drift goes on.
We will now go into the details of each stage.

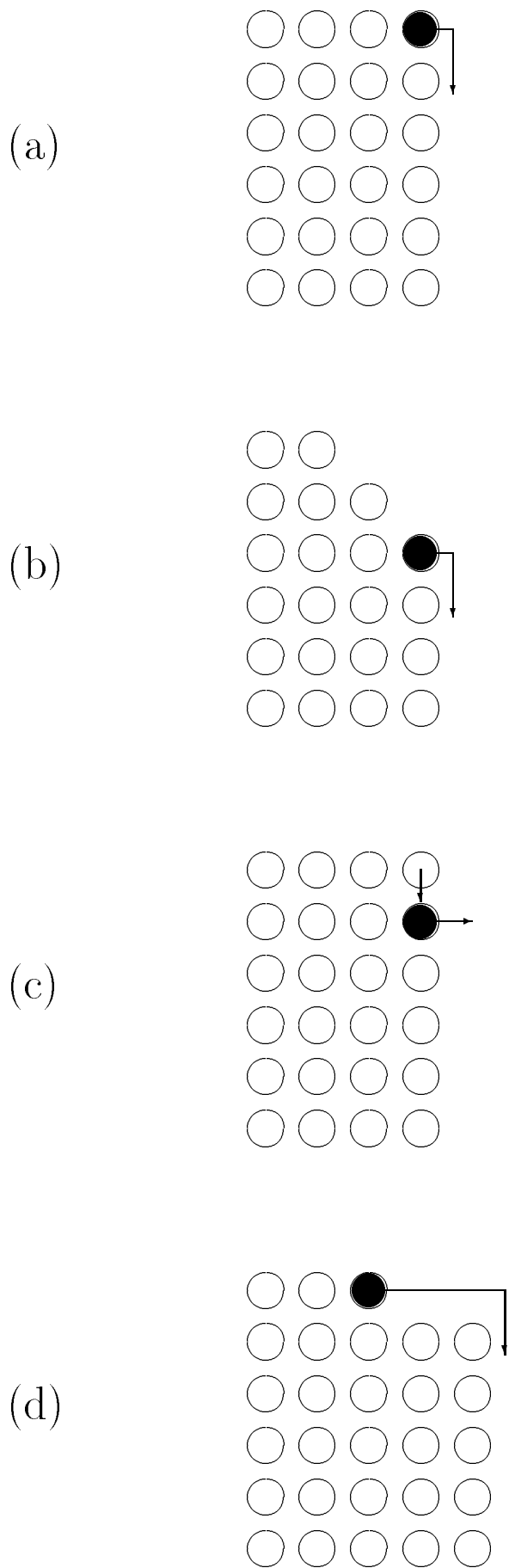
\begin{figure}
\caption{The main mechanisms for detachment that starts a new row:
(a) from a straight corner, (b) from
a kinked corner, (c) from a straight edge
and (d) from a kinked edge.
It is found that (a) has the lowest activation energy
and is thus dominant.} 
\label{fig:isl_escape}
\end{figure}

Consider an island with a straight front edge in the
drift direction as in Fig. \ref{fig:island_drift}(a).
One atom now detaches from one of the corners and
arrives at this front. 
If it detached from a rear corner (with respect to
the drift direction), it will hop along the sides
parallel to the drift under the bias influence,
and arrive at the front.
There are four possible moves in which an atom can leave
the corners, which we index from (a) to (d), as shown in
Figs. \ref{fig:isl_escape}(a-d), respectively.
All these processes actually take place,
but we will now argue that one of them is much more probable,
and thus determines the properties of the drift.
This process is an escape from a straight corner
[Fig. \ref{fig:isl_escape}(a)]. 
By escape here we mean the detachment of an atom from a relatively
bound configuration to the island edge, but not a 
complete detachment from the island. \\
The activation energy for this move is given by:
\begin{eqnarray}
 E_{escape}(a)& = & E_0+E_{NN}+E_{NNN}  \nonumber \\
& = & 0.79 \mbox{eV(model),0.78 eV (EAM)},
\label{eq:escape_a}
\end{eqnarray}
where the first number corresponds to the model 
[Eq. (\ref{eq:model})], and the second number is the
EAM barrier.
After the first of the two hops which compose
the move in Fig. \ref{fig:isl_escape}(a) has taken place,
there are even chances (neglecting the bias)
for that atom to go back to the corner or to actually
escape to the straight edge.
Another possibility is the escape from a straight
edge [Fig. \ref{fig:isl_escape}(b)].
The barrier for such event is 
\begin{eqnarray}
E_{escape}(b)& = & E_0+E_{NN}+2E_{NNN} \nonumber \\
        & =& 0.82 \mbox{eV(model),0.90 eV (EAM)}.
\label{eq:escape_b}
\end{eqnarray}
It may seem at first sight that the extra NNN bond in
comparison to the corner escape may be compensated by the
large number of atoms at this configuration along the
island edges. This is not the case, however, because
an actual escape can happen only one
lattice position off the corners, and not anywhere
along the edges . Otherwise, the
escaped atom is most likely to return to its initial position.
Furthermore, as mentioned before, the barriers obtained
by the model are less accurate at dense environments.
The corresponding EAM barrier for the escape from
straight edge is 0.9 eV.
Another possibility is the escape from a rounded (kinked)
corner [Fig. \ref{fig:isl_escape}(c)].  
The expression obtained from the model for
this barrier is the same as in Eq. (\ref{eq:escape_b}).
The EAM barrier for this move is $E_{escape}(c)=0.83$ eV,
which is slightly higher than that for move (a).
Also, at this configuration several other processes
are much more likely to happen before an escape move
takes place.
The last move (d) is an escape of an atom from an edge kink 
followed by hopping around a corner [Fig. \ref{fig:isl_escape}(d)]. 
The barrier for each of these processes separately 
($E_k$ and $E_c$ respectively) is
lower relative to the previous three:
\begin{eqnarray}
 E_k & = & E_0+2E_{NNN}   \nonumber \\
     & = & \mbox{0.54 eV(model),0.48 eV(EAM)},
\end{eqnarray}
and 
\begin{eqnarray}
 E_c & = & E_0+E_{NNN}   \nonumber \\
     & = & \mbox{0.52 eV(model),0.54 eV(EAM)},
\end{eqnarray}
To understand why this move is relatively unlikely,
it should be noticed that in most cases the atom will
re-attach to the kink rather than hop around the corner.
The barrier for a move from the corner back
towards the kink site is low:
$E_{back}=E_0-E_{NN}+E_{NNN}$ [=0.24 eV(model),0.18 eV(EAM)].
The average number of times the atom has to detach from the kink
until it hops once around the corner is given by 
\begin{equation}
(h_c+h_{back})/h_c\simeq\exp{[(E_c-E_{back})/k_BT]},
\end{equation}
since $h_c \ll h_{back}$ for the relevant temperature range.
The prefactor depends on the distance between the 
kink and the corner.
The effective barrier for the move shown in
Fig. \ref{fig:isl_escape}(d) is
\begin{eqnarray}
E_{escape}(d)& = & E_k+E_c-E_{back}     \nonumber \\
	     & = & E_0+E_{NN}+2E_{NNN}  \nonumber \\
             & = & 0.82 \mbox{eV(model),0.84 eV (EAM)}.
\label{eq:escape_d}
\end{eqnarray}

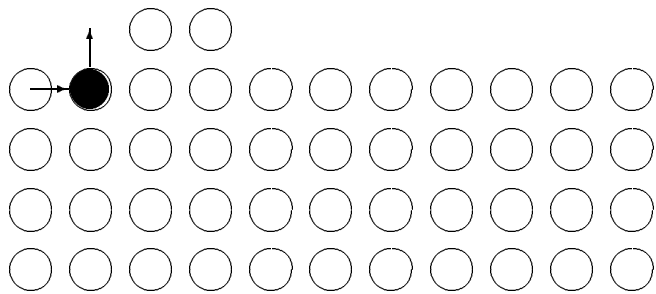
\begin{figure}
\caption{The main mechanism for nucleation of atoms on a new row.
Notice that it can take place only near the corners, and that
the inverse process is identical.}
\label{fig:isl_nucleation}
\end{figure}

The next stage in the island drift process is the nucleation of other
atoms to the atom hopping along the straight facet.
The basic mechanism for this nucleation is shown in
Fig. \ref{fig:isl_nucleation}. 
The energy barrier for this process is
\begin{eqnarray}
 E_{nuc} & = & E_0+3E_{NNN}   \nonumber \\
        & = & \mbox{0.57 eV(model),0.53 eV(EAM)},
\end{eqnarray}
This process must
occur several times in order to create a stable new
row of atoms on the existing facet. Still, the 
time scale of this nucleation process is short compared to
that of the escape process. The ratio between the rates
of these processes is at least
\begin{eqnarray}
\exp{[(E_{escape}-E_{nuc})/k_BT]}\simeq  \nonumber \\ 
\simeq\exp{[(E_{NN}-2E_{NNN})/k_BT]}.
\end{eqnarray}
For Cu, this ratio is 125 at 600 K and more than 15000 at
room temperature.
In the clusters we consider, where a typical 
linear island size is of
10-30 atoms, this is a relatively fast process.

Now, atoms from the edges parallel to the bias direction
will fill-in the kinks created by the atoms that moved to
the new row. This stage 
is necessary to enable the nucleation of the next
new row. 
This is due to the high rate of adatom re-attachment
to kinks.
There are two basic mechanisms for this edge drift which are
shown in 
Fig. \ref{fig:edge_drift}. 
In Fig. \ref{fig:edge_drift}(a) 
a single atom hops along the edge 
in the bias direction.
Eventually this atom will be absorbed in a kink site at 
the island corner.
In Fig. \ref{fig:edge_drift}(b) 
the same happens with an edge vacancy in
the opposite direction.
The barriers for these processes are 
\begin{eqnarray}
 E_{a} & = & E_0-E_{NN}+2E_{NNN}   \nonumber \\
        & = & \mbox{0.27 eV(model),0.25 eV(EAM)},
\end{eqnarray}
and
\begin{eqnarray}
 E_{b} & = & E_0+2E_{NNN}   \nonumber \\
       & = & \mbox{0.54 eV(model),0.48 eV(EAM)},
\end{eqnarray}
The first one is of course much faster, but requires 
an escape move to begin with. 
To estimate the time scale of these processes,
one needs to obtain the average number of 
single hopping events required for one atom/vacancy
that leaves one corner to be absorbed at the opposite one.
It is reasonable to
treat both atoms and vacancies as one dimensional random walkers.
The problem then reduces to a random walk with two absorbing
barriers. The expression obtained from the theory of biased
random walks
(see e.g. Ref. \cite{Srinivasan1974}),
depends on the bias and the length of the walk (i.e. the 
linear size of the island). 
In the range of these two parameters we deal with,   
the results are 26-200 events. Again, this is a short
time scale compared to that of escape events.
In summary, the drift of islands can be divided into three
stages with three different time scales.
The velocity of the drift is determined by the slowest of these,
namely the escape event.
We come to the conclusion that
the escape process (a) is the rate limiting
step of island drift. 
The effective barrier
for the drift is thus
\begin{eqnarray}
E_{B}(\mbox{island drift}) = E_{escape}(a) \nonumber \\
	= E_0+E_{NN}+E_{NNN} \nonumber \\
                     =  0.79 \mbox{eV(model),0.78 eV (EAM)} .
\label{eq:isl_drift_barr}
\end{eqnarray}

\subsection{Void Drift}

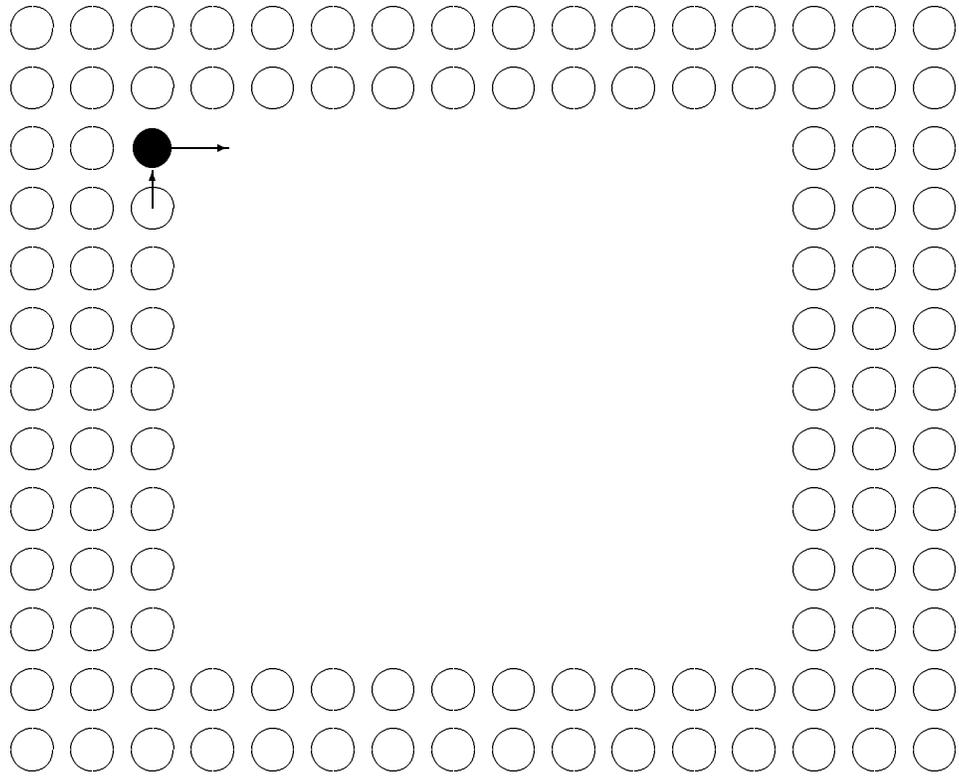
\begin{figure}
\caption{The atomic process that starts a new vacancy
row in a void drift. An atom is detached from the 
straight corner . Another atom may fill its place
and create a stable dimer.}
\label{fig:void_escape}
\end{figure}

Basically, the three stages that exist in the island drift
process, are found for vacancy clusters as well.
The atomic details however, reveal important differences.
The most essential difference is in the equivalent process to the 
corner escape move. Consider a void with a straight
corner (Fig. \ref{fig:void_escape}).
The barrier for this detachment process is
\begin{eqnarray}
 E_{detach} & = & E_0+3E_{NNN}   \nonumber \\
       & = & \mbox{0.57 eV(model),0.53 eV(EAM)}.
\end{eqnarray}
In most cases however, the detached atom does 
not escape from the corner, but goes back
to its initial position. There are several possible moves
in the nearby environment that may prevent the detached
atom from moving back.
The one with the lowest barrier is 
the replacement move shown
in Fig. \ref{fig:void_escape}. The average number of
times a detachment event should occur before such move takes
place, is very well approximated by the ratio of
the rates 
\begin{equation}
h_{back}/h_{replace}=\exp{[(E_{replace}-E_{back})/k_BT]}.
\end{equation}
The effective barrier obtained for an escape from 
the corner is
\begin{eqnarray}
E_{escape} & = & E_{detach}+E_{replace}-E_{back} \nonumber \\
	   & = & E_0+E_{NN}+3E_{NNN}  \nonumber \\
           & = & 0.84 \mbox{eV(model),0.73 eV (EAM)}.
\label{eq:island_barr}
\end{eqnarray}

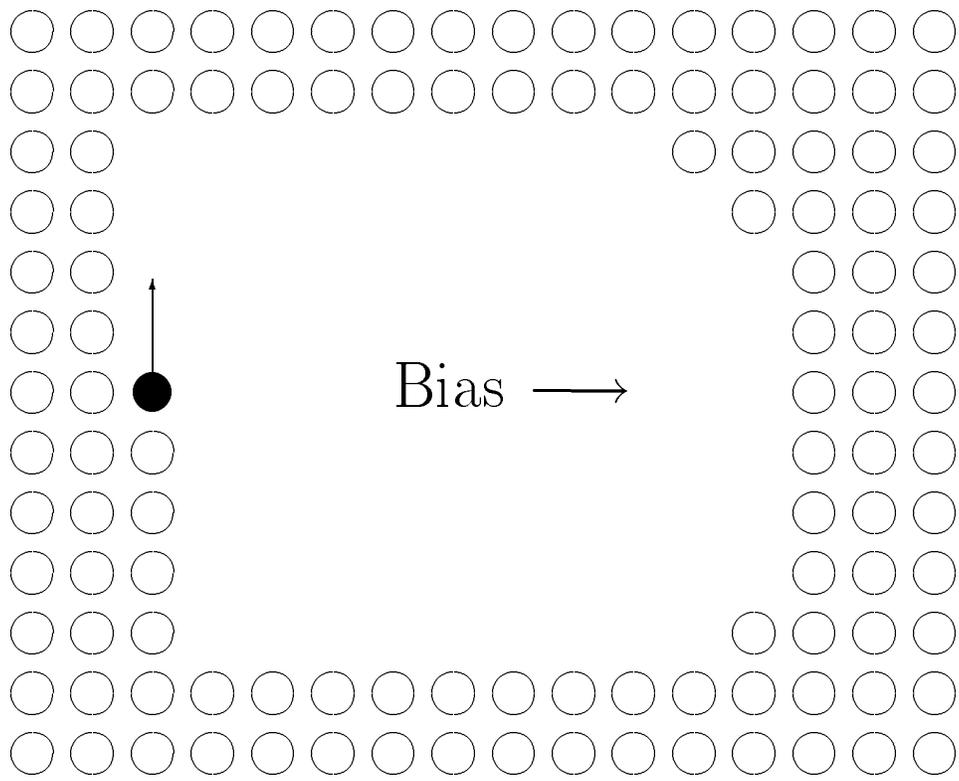
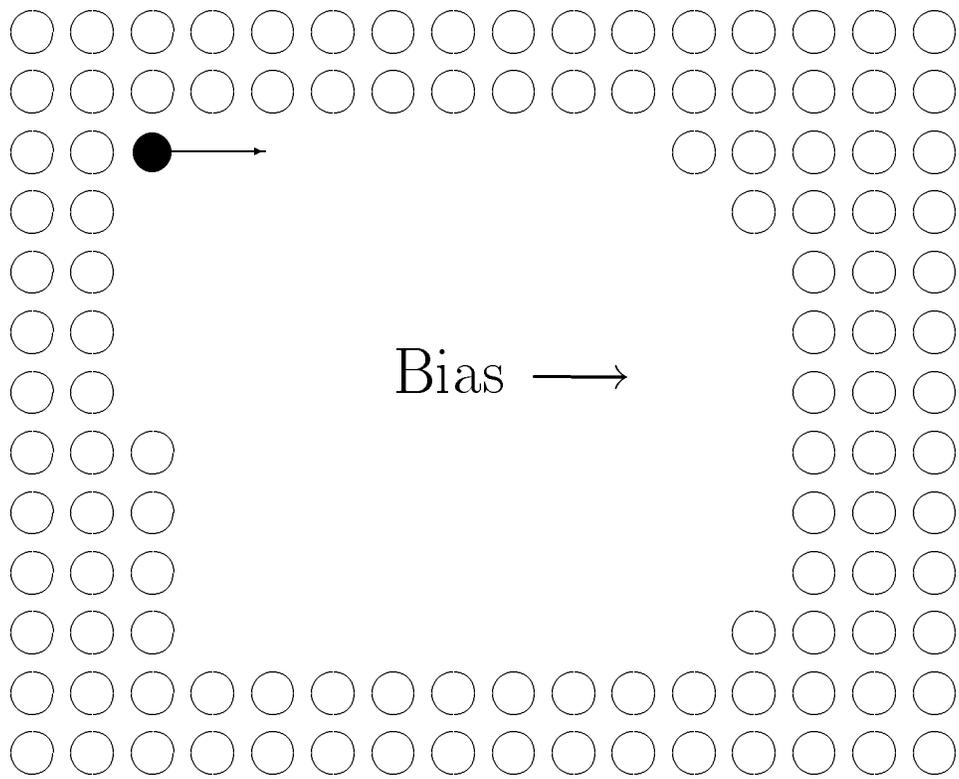
\begin{figure}
\caption{Nucleation of vacancies at the new vacancy row is
actually composed of atom detachment from an edge kink (a),
hopping to the corner of the void (b), and then to the opposite 
corner.}
\label{fig:void_nucleation}
\end{figure}

The nucleation of other vacancies on the void edge is
created mainly by the detachment of atoms from an edge
kink and their diffusion along the edge
(Fig. \ref{fig:void_nucleation}).
The bottleneck of this process is the move across the 
corner of the void, shown in  
Fig. \ref{fig:void_nucleation}(b).
The barrier for leaving the corner is
$E_c=E_0+3E_{NNN}$ [=0.57 eV(model),0.53 eV(EAM)].
\begin{eqnarray}
 E_{c} & = & E_0+3E_{NNN}   \nonumber \\
        & = & \mbox{0.57 eV(model),0.53 eV(EAM)},
\end{eqnarray}
In contrast to islands, void corners are ``attractive''.
Since the barrier for leaving the corner is comparable
to the barrier for leaving edge kink, in many cases there
will be no accumulation of atoms near the corner, and 
we will not get the rounded corners as in islands.
In other words, the stages of nucleation and corner
straightening are combined in voids.
In order to compare the time scales of the stages, we need
to estimate how many corner detachment events are required for
one atom to reach the opposite edge of the void.
(The hopping on the edge itself is again very fast and
its actual time is negligible.)
Using again the approximation of a one dimensional
random walk, we obtain an estimation of $\approx10$
such events for voids of linear size 10. 
Since all atoms on the facet must
go through the corner, at least 100 events are
required for the nucleation of a complete row.
The time scale of the nucleation stage is thus short
compared to that of the escape event at low 
temperatures. 
Under these conditions the effective barrier for void drift is
\begin{eqnarray}
E_{B}(\mbox{void drift}) = E_{escape} = E_0+E_{NN}+3E_{NNN} \nonumber \\
		= 0.84 \mbox{eV(model),0.73 eV (EAM)}.
\label{eq:void_drift_barr}
\end{eqnarray}
At higher temperatures, however, depending on the 
size of the cluster and the bias, the two time
scales may become comparable.  

\subsection{The Drift Velocity}

For both islands and voids we identified
the rate limiting processes and their energy
barriers. These barriers are the effective activation
energies of the entire drift process. We thus expect
the drift velocities of islands and voids to scale like
\begin{equation}
v_{drift}(\mbox{island})\sim \exp(-\frac{E_0+E_{NN}+E_{NNN}}{k_BT})
\label{eq:island_drift}
\end{equation}
and
\begin{equation}
v_{drift}(\mbox{void})\sim \exp(-\frac{E_0+E_{NN}+3E_{NNN}}{k_BT})
\label{eq:void_drift}
\end{equation}
respectively.
It is convenient to measure the velocity of the 
cluster center of mass in units of lattice sites per
second. 
We expect from Eq. (\ref{eq:velocity}) that the drift velocity
is linearly proportional to the  bias magnitude.
It is also
proportional to the probability of
an escaped atom to form a new row
rather than go back.
This probability
cannot be deduced from simple arguments, since it
generally involves a random walk with
moving boundaries and many particles.
In general it may depend on the cluster size and
the temperature.
For larger clusters, more atoms 
need to nucleate to complete a new front row. 
It also takes longer to fill the kink site
created by each newly nucleated atom, since the sides parallel
to the bias component are now longer.
We thus expect that the prefactor 
decreases as the cluster size
increases, but cannot provide a specific functional form.

\section{Simulations and Results}

In the simulations reported below
we used the continuous time kinetic MC technique 
\cite{Bortz1975,Voter1986,Fichthorn1991,Lu1991,Clarke1991,Kang1994,Barkema1994}.
This technique is particularly suitable for the simulation
of non-equilibrium processes, keeping track of 
the physical time in a realistic manner.
During the kinetic MC simulation, the next move is
selected randomly from the list of all possible moves
at the given time with the appropriate weights. 
The time is advanced after each move
according to the inverse of the sum of all rates.
All the processes allowed by the model are
incorporated in the simulations with the 
appropriate energy barriers.
In the simulation results presented
below we used barriers for Cu/Cu(001)
obtained by EAM \cite{Biham1998}. 

We have performed systematic simulations on single isolated
islands and voids. 
The initial island and void configurations were chosen to be of
a square shape from the reasons already discussed. 
The morphology and location of the clusters were followed
for different temperatures, bias directions, and bias magnitudes. This was
repeated for different sizes of the islands/voids.
As expected, islands drift in the direction of the bias, while vacancy 
clusters drift in the opposite direction.
In the temperature range considered here ($200-500K$), both islands 
and voids drift as a whole, since the activation energy for
a complete detachment of atoms or vacancies 
is high relative to the processes discussed above.
Furthermore, even if an atom/vacancy is 
detached from the cluster, the bias is not strong enough for a substantial
drift, and it will re-attach after few moves. 

In the simulations we follow the displacement of
the center of mass of the clusters as a function of
the physical time.
It was found that under the EM bias, clusters drift 
on the average at
a constant velocity.  
The dependence of this velocity on the cluster size,
bias magnitude and the temperature is shown below.

\subsection{The Effect of Bias Magnitude on the Drift Velocity.}

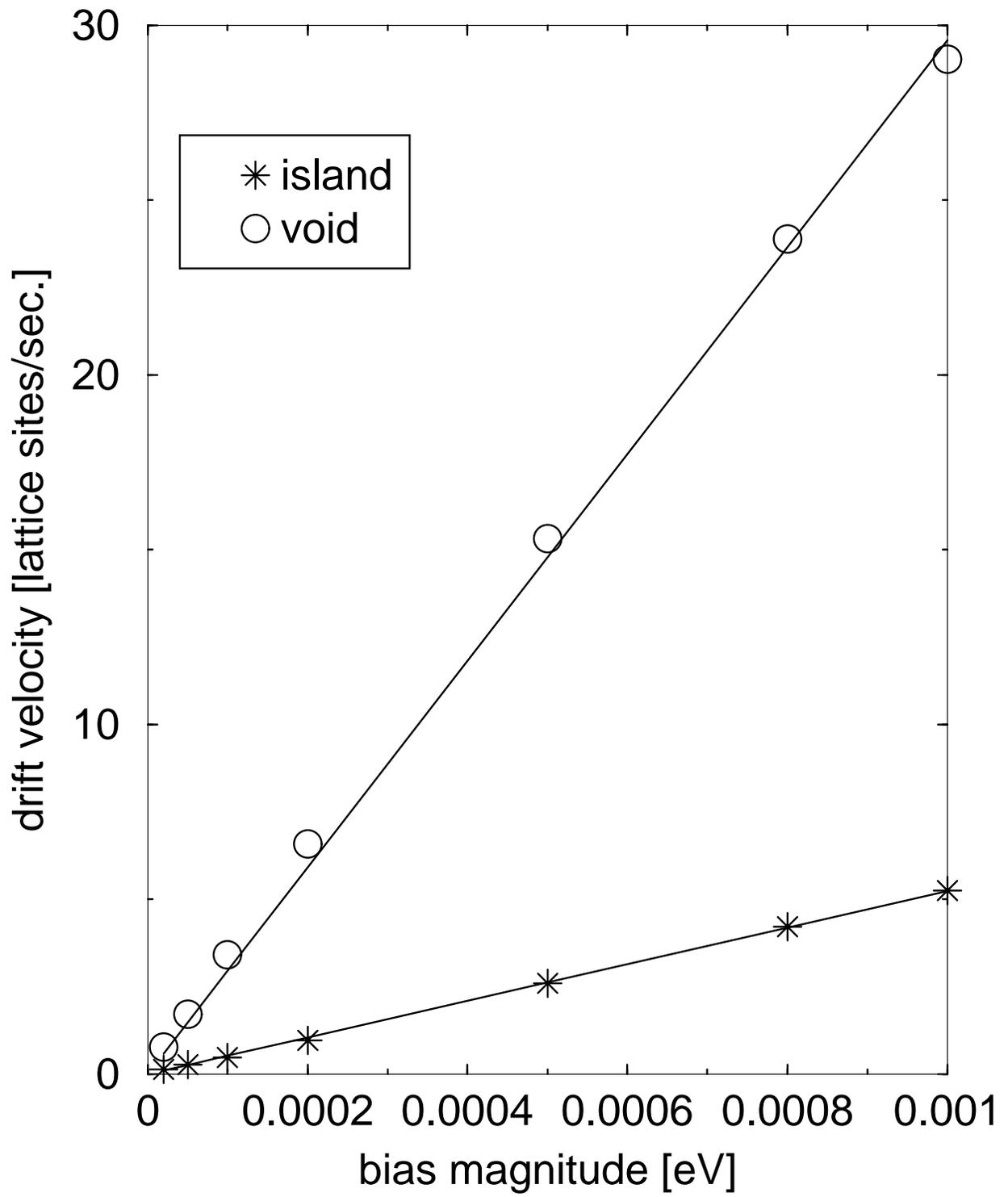
\begin{figure}
\caption{Drift velocity as a function of bias magnitude
for $10\times10$ clusters.  
Stars are voids at temperature of 350K and
Circles are islands at 400K.
Error bars are about the size 
of the symbols.}
\label{drift:vs:bias}
\end{figure}

We found that the drift velocity of islands 
depends linearly on the bias as can be seen
from Fig. \ref{drift:vs:bias}. 
This dependence is expected from Eq. (\ref{eq:velocity}),  
and was confirmed by simulations for bias magnitudes
in the range $\Delta\approx10^{-3}$--$10^{-5}$ eV. 
In addition we confirmed by simulations that
the effect of bias direction is given by
$v_x\propto\Delta\cos(\phi)$ and  $v_y\propto\Delta\sin(\phi)$,
where $v_x$ and $v_y$ are the $x$ and $y$
components of the drift velocity, and $\Delta$ is
the bias magnitude applied in  
an angle $\phi$ relative to the $x$ axis.  

\subsection{Drift Velocity as a Function of the Temperature}

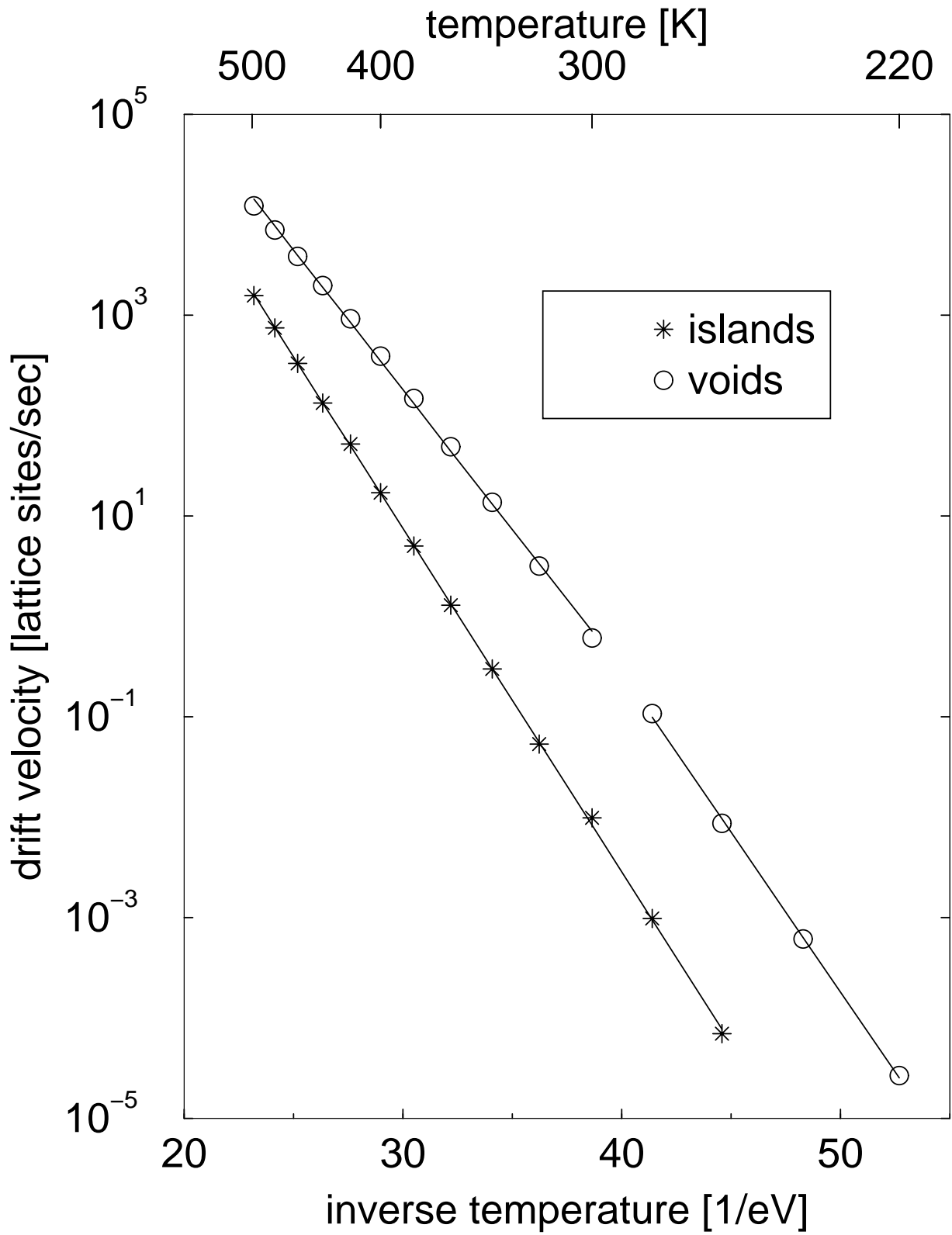
\begin{figure}
\caption{Drift velocity as a function of $T^{-1}$ 
on semi-logarithmic scale, for 
$10\times10$  island (stars) and void (circle) 
with bias of 0.003 eV.
Solid lines are a fit to an exponential form
$A\cdot\exp({\frac{-Q}{k_BT}})$.
The temperature range is 300-400K and the error bars 
are smaller than the symbols.
The slopes obtained from the fit are $Q=0.78\pm2$ eV
for islands. For voids at $T<300$ K $Q=0.73\pm2$ eV.
These values are in agreement with the barriers of
the rate limiting steps discussed in the text.
}
\label{drift:vs:temp}
\end{figure}

The dependence of the drift velocity on the temperature is shown in
Fig. \ref{drift:vs:temp}.
The activation energy for island diffusion, is 
found by the best exponential fit to be \\ 
$E_B$(island drift)$=0.78\pm0.02$ eV. 
This coincides with the energy barrier for the escape move,
which we identified as the rate limiting step
in the analysis above.
For voids, we find that for temperatures 
in the range $220K<T<300K$ there is 
a good fit to the predicted value: \\
$E_B$(void drift)$=0.73\pm0.02$.
At higher temperatures there is a small deviation from
this value towards a lower value.
This may indicate that the probability of an 
atom that escaped from a void corner to be 
re-attached rather than start a new vacancy row,
slightly depends on the temperature.

\subsection{Dependence of The Drift Velocity on Cluster Size}

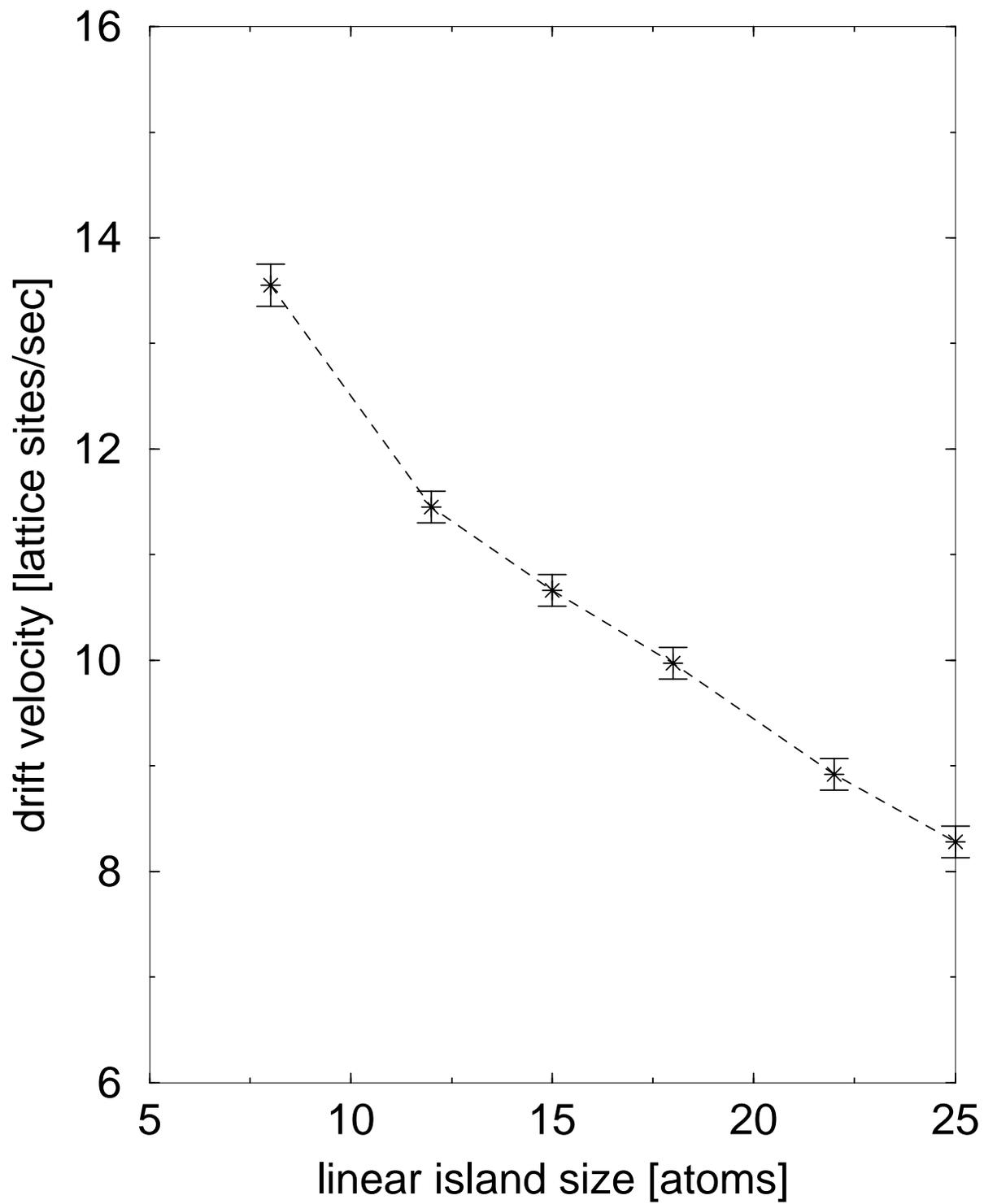
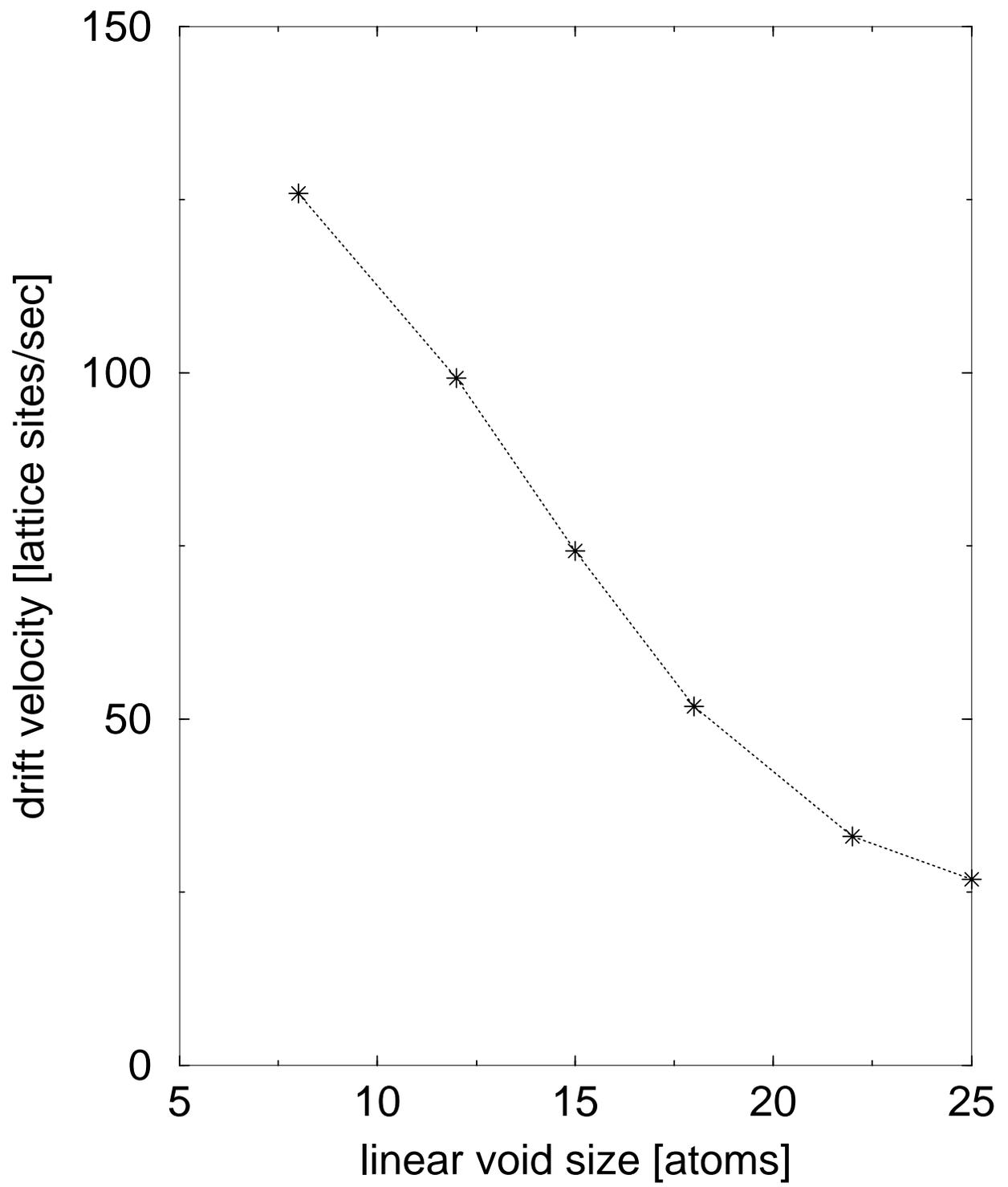
\begin{figure}
\caption{Drift velocity as a function of cluster size for
(a) islands at bias of 0.002 eV  
and (b) voids at bias of 0.0005 eV.
The temperature is 400K. 
The error bars in (a) correspond
to one standard deviation of the runs, while in (b) they are
smaller than the symbols.}
\label{drift:vs:size}
\end{figure}

Fig. \ref{drift:vs:size} shows the dependence of the drift 
velocity on the linear size of the cluster.
As could be expected, there is a 
monotonic decrease in drift velocity. 
The quantitative details, however, are not completely concluded. 

\section{Discussion and Summary}

In this paper we have studied the mechanisms 
of EM-driven diffusion of single islands and voids
on the Cu(001) surface.
We found that the drift velocity of a cluster of a given
linear size depends on the bias $\Delta$ and the temperature $T$
according to
\begin{equation}
\label{eq:drift}
v(T,\Delta)=A\cdot\frac{\Delta}{k_BT}\nu\cdot\exp{(-Q/k_BT)},
\end{equation}
\noindent
where Q is the activation energy of  the rate limiting
step of the drift process and A is a constant.
The reciprocal dependence on the temperature 
could not be inspected from the simulation results,
since it is
much weaker than the exponential dependence. 
It is deduced, however, from the 
discussion following Eq. (\ref{eq:velocity}), 
and from dimensional analysis.
The activation energy Q is
given by Eq. (\ref{eq:isl_drift_barr}) for islands 
and by Eq. (\ref{eq:void_drift_barr}) for voids.

The value of the constant A 
is proportional to the 
probability of an escaped atom to form solid
nucleation and establish a new front row of atoms.
As we mentioned before, the atomic diffusion processes
that determine this probability are complicated. 
The simulations, however, yield a value of
$A\sim5-20$ for islands, and
$A\sim40-80$ for voids, depending on the cluster size,
provided that the velocity is measured in units of 
lattice sites per second.

It was found 
\cite{Mehl1999} that most FCC(001) metal surfaces share
the same qualitative features of
the different hopping mechanisms.
The hopping energy barriers in these systems can
be parameterized by models similar to the one we used
here, with the specific parameters appropriate for 
each metal. 
The analysis of the atomic processes involved in
the drift of islands and voids,
and their typical time scales should be
thus valid for these FCC(001) metal surfaces as well.
Several simulations performed
for Ag/Ag(001) diffusion, indicate that this is
indeed the case.
 
Extrapolation of the results to other regimes are
reasonable as long as the relations between the time scales
remain as in the above discussion. Larger clusters 
or weaker bias may be considered as long as
the nucleation and corner straightening stages are still
faster relative to the escape move. For islands at room
temperatures, for example, this means clusters
of up to $\sim10^4$ atoms at bias of 0.001 eV, or
bias down to $\sim10^{-6}$ eV for clusters of $\sim100$ atoms. 
The linear dependence of the drift velocity on the bias 
makes the extrapolation down to a more realistic bias
values straightforward.

In order to test our predictions 
experimentally one needs to prepare Cu(001) surfaces with islands of
the desired size distribution.
Then one needs to drive current along the surface
at high current densities.
Using scanning tunneling microscope (STM)
one will be able to measure the drift velocities of different
islands and obtain their dependence on the island size, bias and temperature.
Related experimental work has been done on diffusion 
of islands on Cu(001) with no bias 
\cite{Pai1997}
and the dependence of the
diffusion coefficient on island size was obtained.
To our knowledge, no such experiments on single 
crystal metal surfaces have been done
under EM conditions.
It seems that a serious difficulty in performing 
such experiments on single
crystal samples may be to reach the 
high current densities required to obtain 
a fast enough drift, due to the considerable width of such samples.

\acknowledgements

We thank J. Krug for helpful discussions.
We would like to acknowledge support from 
the Intel-Israel college relations committee
during the initial stages of this work, and from
the Israeli Ministry of Science and Technology
during the final stages.

\begin{figure}
\caption{Classification of all possible local environments 
of a hopping atom including seven adjacent sites. 
Each site can
be either occupied or unoccupied, 
giving rise to $2^{7}=128$ local 
environments. 
Sites 1, 3 and 5 are nearest neighbors of the original 
site while sites 1, 2, 5 and 6 are adjacent to the bridge site 
that the atom has to pass.}
\label{fig:env} 
\end{figure}

\begin{figure}
\caption{The main stages in cluster drift starting from
a straight facet (a). 
An atom is detached from the corner and
starts a new row (b). Other atoms nucleate to the first
atom and complete the row (c). Atoms then drift along the 
edges and retain the straight facet . The arrow 
on the grid indicates the initial position
of the lattice front.}
\label{fig:island_drift}
\end{figure}

\begin{figure}
\caption{The main mechanisms for atom drift along island edges.
(a) A single atom hops on a straight edge; and (b) an edge
vacancy drifts in the opposite direction. The arrow 
indicates the bias direction.}
\label{fig:edge_drift}
\end{figure}	

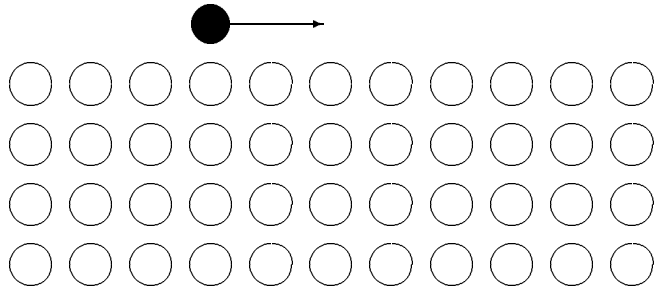
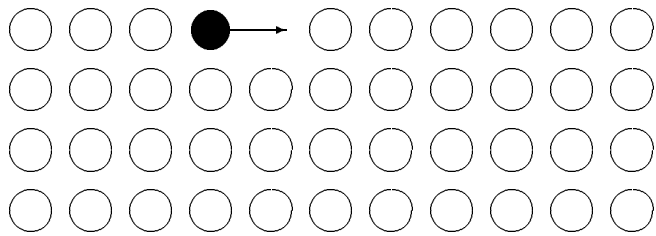
\end{document}